\documentstyle[12pt]{article}
\begin{document}

\title {SYMMETRIES AND STRING FIELD THEORY IN D=2}
\author {Michio Kaku}
\date {City College of the CUNY, N.Y., N.Y. 10031}
\maketitle

\begin {center}
\Large {\bf Abstract}

\end {center}

Prof. Wigner was well-known for his contributions to group theory
and symmetries, especially the representations of the Poincar\'{e} group.
Similarly, symmetry also plays a decisive role in string theory.
In fact, because of these symmetries,
in two dimensions the theory is, remarkably enough,
exactly soluble, giving us the first
non-perturbative information from string theory.

There are two ways in which to
formulate 2D string theory, either as a matrix model
or as a Liouville theory [1-4].
Matrix models are enormously powerful, and the emergence
of a new symmetry, called $w (\infty)$, helps to explain why the
model is exactly soluble.
Its generators obey:

\begin {equation}
[ Q _ { j m  } , Q _ { j ' m ' } ] =
( j m ' - j ' m ) Q _ { j + j ' -1 , m + m' }
\end {equation}
where $j = 0 , 1/2 , 1 , 3/2 , ... $ and $m = - j , ... + j$.

However, in matrix models the
string degrees of freedom have been completely removed, and
hence the field theoretic
origin of $w (\infty)$ is rather obscure. Furthermore,
the generators of $w(\infty)$ involve the so-called \lq\lq discrete states,"
which are unusual physical states defined only at discrete momenta.

Liouville theory, by contrast, has the opposite problem: it
manifestly has
all the string degrees of
freedom intact, but the theory is notoriously difficult to solve, even at the
free level.
The purpose of this paper is to show how to
reformulate Liouville theory as a
second quantized field theory, so that $w(\infty)$
and the discrete states emerge
naturally.

To re-write Liouville theory (for $\mu = 0)$
as a field theory, we will generlize the
26 dimensional
non-polynomial closed string field theory, recently proposed
by the author [5] and also by the Kyoto and
MIT groups [6,7], which solves the long-standing problem of
a covariant, closed string field theory.

Using this non-polynomial theory in two dimensions,
we will show that the
origin of $w(\infty)$ comes from the gauge group of the
non-polynomial
action.
Furthermore, we will show how to reproduce the
standard shifted Shapiro-Virasoro four-point
amplitude from the non-polynomial Liouville
field theory.
This gives a cohesive, integrated description
of the many seemingly divergent results found in this
rapidly growing field.

Further details of this paper can be found in ref. [8].

\section {Introduction}

Usually, when we have a symmetry of a theory, we start with a Lie
algebra $ [ \tau _ a , \tau _ b ] = f _ { abc }\tau _ c $. Then we
let the second quantized field $A _ \mu ^ a$
transform in some representation of the group, so that:
$\delta A _ \mu ^ a = f _ { abc } A _ \mu ^ b \Lambda ^ c +...
$. And lastly, we write down an action ${\cal L}$ which is
an invariant under this variation. We find, for example, that the
three-vertex is proportional to the structure constant
of the  group $f _ { abc }$.

However, $w(\infty)$ emerges from matrix models in a highly
unorthodox fashion, largely because the string degrees of
freedom have been completely eliminated.

The goal of this paper is to reformulate closed string Liouville theory
as a second quantized field theory, so that $ w ( \infty)$ emerges in
the usual way. We will show that the second quantized fields
transform under a certain representation of the group, that
the action is an invariant of the group, and that the three-vertex
is proportional to the structure constant of the group.

The action of the 26 dimensional string theory is given by the
non-polynomial closed string action, first written down by
the author [5] and the Kyoto and MIT groups [6,7]:

\begin {equation}
{\cal L} = \langle \Phi | Q | \Phi \rangle +
\sum _ { n = 3 } ^ \infty \alpha _ n
\langle \Phi ^ n \rangle
\end {equation}
where $Q = Q _ 0 ( b _ 0 - \bar b _ 0 )$, $Q _ 0$
is the usual BRST operator, and
where the field $\Phi (X) $ transforms as:

\begin {equation}
\delta | \Phi \rangle =
| Q \Lambda \rangle +
\sum _ { n = 1 } ^ \infty \beta _ n | \Phi ^ n \Lambda \rangle
\end {equation}

In ref. 8, we have genalized this action to describe
two dimensional string theory.
In this paper, we will show that:

(a) the vertices are BRST invariant

(b) the action reproduces the shifted Shapiro-Virasoro
amplitude

(c) the three-vertex is proportional to the structure constant of
$ w ( \infty)$.

\section {BRST Invariance of Vertices}

We first wish to show that the three-vertex is BRST invariant, i.e.

\begin {equation}
\sum _ { i = 1 } ^ 3 Q _ i | V _ 3 \rangle = 0
\end {equation}
Naively, this calculation appears to be trivial, since
the vertex function
simply represents a delta function across three overlapping strings.
Hence, we expect that the three contributions to $Q$ cancel exactly.
However, this calculation is actually
rather delicate, since there are potentially
anomalous contributions at the joining points.
To resolve this issue, we must use point-splitting, pioneered
in refs. [9,10].

We wish to make a conformal map from the multi-sheeted,
three-string, world-sheet configuration
in the $\rho$-plane to the flat, complex $z$-plane.
Fortunately, the conformal map for the $N$-point function to the
complex $z$-plane is known. The map is given by [5]:

\begin {equation}
{ d \rho (  z )
\over d z }
= C { \prod _ { i = 1 } ^ { N - 2 }
\sqrt  { ( z - z _ i ) ( z -  \tilde z _ i ) }
\over
\prod _ { i = 1 } ^ N  ( z - \gamma _ i ) }
\end {equation}
where $\gamma _ i$ correspond to the points at infinity (the external
lines) and the $z _ i, \tilde z _ i $ correspond to the
$i$th vertex, which are interior points.
Then eq. (4) can be transformed to the $z$-plane, giving us:

\begin {equation}
\oint _ { C _ 1 +C _ 2 + C _ 3 }
{dz \over 2 \pi i }
c( z ) \left \{
- { 1 \over 2 } \left ( { dz ' \over  dz } \right)
\partial X _ \mu ( z ' ) \partial X ^ \mu ( z )
+
\left ( { dz ' \over dz } \right ) ^ 2
{ dc \over dz } b ( z ' )
+ { Q \over 2 }
\partial ^ 2 \phi ( z ) \right \}
| V _ 3 \Big \rangle
\end {equation}
where $z'$ is infinitesimally close to $z$,
where $\mu$ ranges over the $D$ dimensional string modes as well
as the $\phi$ mode,
where $b$ and $c$ are the usual reparametrization ghosts,
and $C _ i$ are infinitesimally small
curves in the $z$-plane which encircle
the joining point $z _ 0$.

The major complication to this calculation
is that the Liouville $\phi$ field does
not transform as a scalar. Instead, it transforms as:

\begin {equation}
\phi ( \rho ) \rightarrow \phi (  z ) +
{ Q \over 2 } \log \Big | { d z \over d \rho } \Big | ^ 2
\end {equation}
This means that the energy-momentum tensor $T$ transforms as:

\begin {equation}
T _ { \rho  \rho } \rightarrow
\left ( { d z \over d \rho } \right) ^ 2 T _ { zz }
+ \left( { Q \over 2 } \right ) ^ 2 S
\end {equation}
where $S$ is the Schwartzian, given by
$( z ''' / z ' ) - ( 3/2 ) ( z '' / z ' ) ^ 2 $.
This complicates the calculation considerably, since it means that there is
subtle insertion factor
located at delta-function
curvature singularities in the vertex function.
These add non-trivial $\phi$
contributions to the calculation.

The final calculation is rather long [8], so we only summarize the
result:

\begin {equation}
\left \{ - p c ( z _ 0 ) \left [
{ D \over 24 } - { 13 \over 12 } + { 1 \over 24 }
+ { 1 \over 8 } Q ^ 2 \right ]
- { d c ( z _ 0 ) \over d z }
\left [ { 5 D \over  96 }
- { 65 \over 48 }
+ { 5 \over 96 }
+ { 5 \over 32 } Q ^ 2 \right ]
\right \} | V _ 3 \rangle
\end {equation}
which cancels if:

\begin {equation}
D - 26 + 1 + 3 Q ^ 2 = 0
\end {equation}
which is precisely the consistency equation for Liouville theory
in $D$ dimensions.
Thus, the vertex is BRST invariant.

\section {Shifted Shapiro-Virasoro Amplitude}

The next major test of the theory is whether it reproduces the
shifted Shapiro-Virasoro amplitude. This calculation is
highly non-trivial, since the conformal map between the multi-sheeted
string-scattering Riemann sheet to the complex plane is
very involved. Unlike the light cone theory, or even Witten's
open string theory, the non-polynomial theory yields very
complicated conformal maps.

Fortunately, for the four-point function, all conformal maps
are known exactly, in terms of elliptic functions, and the
calculation can be performed (see ref. 11).

For the four point function, the map in eq. (5) can be integrated
exactly, giving:

\begin {eqnarray}
\rho & = &
\sum _ { i = 1 } ^ 4
{ g N A _ i \over
a _ 1 + b _ 1 g _ 1 - g _ 1 \gamma _ i }
{ \omega _ i - g \over 1 + \omega _ i ^ 2 }
\nonumber \\
&& \times
\left [ \omega _ i ^ 2
\Pi ( \phi , 1 + \omega _ i ^ 2  , k ' )
+ \omega _ i ^ 2 ( \omega _ i ^ 2 + 1 )
 f _i \right ]
\end {eqnarray}
where $\Pi$ is a third elliptic function,
$ z = i a _ i + b _ i$ and $\tilde z _ i
= - i a _ 1 + b _ 1 $ for complex $a _ i$ and $b _ i$, and:
\begin {equation}
A _ i = {  \left [ ( \gamma _ i - b _ 1 ) ^ 2 + a _ 1 ^ 2 \right ]
\left [ ( \gamma _ i - b _ 2 ) ^ 2 + a _ 2 ^ 2  \right ]
\over
\prod _ { j = 1 , j \neq i } ^ 4
( \gamma _ i - \gamma _ j ) }
\end {equation}

\begin {eqnarray}
\omega _ i &=& { a _ 1 + b _ 1 g _ 1 - \gamma _ i g _ 1
\over b _ 1 - a _ 1 g _ 1 - \gamma _ i }
\nonumber \\
f _ i &=&
{ 1 \over 2 } ( 1 + \omega _ i ^ 2 ) ^ { - 1 /2 }
( k ^ 2 + \omega _ i ^ 2 ) ^ { - 1/2 }
\nonumber \\
&& \times \ln
{ ( k ^ 2 + \omega _ i ^ 2 ) ^ { 1/2 }
- ( 1 - \omega _ i ^ 2 ) ^ { 1/2 }
{ \rm dn } u
\over
( k ^ 2 + \omega _ i ^ 2 ) ^ { 1/2 }
+ ( 1 + \omega _ i ^ 2 ) ^ { 1/2 }
{ \rm dn } u }
\nonumber \\
\phi & = & {\rm arc tan}
\left( { y - b _ 2 + a _ 1 g _ 1
\over a _ 1 + g _ 1 b _ 1 - g _ 1 y } \right )
\end {eqnarray}
where:
$A ^ 2 = ( b _ 1 + b _ 2 ) ^ 2 +
( a _ 1 + a _ 2 ) ^ 2
$, $ B ^2  =
( b _ 1 - b _ 2 ) ^ 2 + ( a _ 1 - a _ 2 ) ^ 2$,
$ g _ 1 ^ 2 = [4 a _ 1 ^ 2 - ( A - B) ^ 2 ]
[ ( A + B ) ^ 2 - 4 a _ 1 ^ 2 ] $,
$ g = 2 / ( A + B )$, $ y _ 1 = b _ 1 - a _ 1 g _ 1$,
$ { k '} ^ 2 = 1 - k ^ 2 = 4AB / ( A + B ) ^ 2 $,
$ u =  {\rm  dn} ^ { -1 } ( 1 - { k '}  ^ 2  {\rm sin} ^ 2 \phi )$

This explicit conformal map allows us to calculate the four-point
amplitude.
We first write the amplitude in the $\rho$ plane, and then make a
conformal map to the $z$-plane.   Let the modular parameter be
$\hat \tau = \tau + i \theta $, where $\tau$ is the distance
between the splitting strings, and $\theta $ is the relative
rotation. Let $\gamma _ 1 =
( 0 , \hat x , 1 , \infty)$. Then, with a fair amount of work,
one can find the Jacobian from $\hat \tau$ to $\hat x$:

\begin {equation}
{ d \hat \tau \over d \hat x }
=
{ \pi C \over
2 K( k ) g \hat x ( 1 - \hat x )
( \gamma _ 1 - \gamma _ 3 )( \gamma _ 2 - \gamma _ 4)}
\end {equation}

Then the four point amplitude can be written as:

\begin {eqnarray}
A _ 4 &=& \langle V _ 3 | { b _ 0 \bar b _ 0 \over
L _ 0 + \bar L _ 0 - 2 } | V _ 3 \rangle
\nonumber \\
& = &
\int d \tau \int d \theta
A _ G \Big \langle \prod _ { i = 1 } ^ 4
c ( \gamma  _ i  ) \bar c ( \gamma  _ i  ) V ( \gamma  _ i  )
\Big \rangle
\end {eqnarray}
where:

\begin {eqnarray}
A _ G &=& \int { d z \over 2 \pi i }
{ d z \over dw }
{ \prod _ { i < i } ( \gamma _ i - \gamma _ j )
\over \prod _ { j = 1 } ^ 4
( z - \gamma _ j ) }
\noindent \\
& = &
2 { g \over \pi c }  \hat x ( 1 - \hat x ) K ( k )
( \gamma _ 1 - \gamma _ 3 ) ^ 3 ( \gamma _ 2 - \gamma _ 4 ) ^ 3
\end {eqnarray}

Putting everything together, we finally find:

\begin {equation}
A _ 4 = \int d ^ 2 \hat x \Big |
\hat x ^ { 2 p _ i \cdot p _ j }
( 1 - \hat x ) ^ { 2 p _ 2 \cdot p _ 3 } \Big | ^ 2
\end {equation}
for shifted momenta $p$. This is
precisely the shifted Shapiro-Virasoro amplitude,
as expected.

\section {Origin of $w ( \infty)$}

Finally, we wish to show that
$w ( \infty)$ emerge in a natural way
in the second quantized Liouville theory,
i.e. the structure constants of $w( \infty)$
are proportional to the three-string vertex function.

First, notice that the field $|\Phi \rangle$ can be
decomposed into a massless tachyon, the discrete
states $|j,m \rangle$ (labeled by $SU(2)$ quantum numbers $j,m$),
and BRST trivial states:

\begin {equation}
|\Phi \rangle =
\varphi | 0 \rangle +
\sum _ { j,m } \psi _ { j,m}
| j,m \rangle + ...
\end {equation}
where $...$ represent the BRST trivial states.

Now take the matrix element of
three strings:
$\langle \Phi | \langle \Phi | \langle \Phi |
V _ 3 \rangle$. When $\langle \Phi |$ is physical and
on-shell,
only the tachyon and discrete states survive.
To compute the three-vertex function between three
discrete states, we will make a conformal transformation
from the three-vertex $\rho$ complex plane to the
complex $z$-plane. Under a complex transformation,
$| j,m \rangle$ remains the same, i.e.
$ \Omega | j, m \rangle = | j,m \rangle$ since
$| j,m \rangle$ satisfies the Virasoro conditions.
Let us therefore insert $\Omega \Omega ^ { -1}$
inside the three-string vertex function.
The $| j, m \rangle$ states remain the same, but the
vertex function $ | V _ 3 \rangle _ { \rho }$
transforms onto $| V _ 3 \rangle _ z$, which is defined on the
complex $z$-plane

Thus, it is a simple matter to show (for the holomorphic part),
that:
\begin {eqnarray}
\langle j _ 1 , m _ 1  | \langle j _ 2 , m _ 2 |
\langle j _ 3 , m _ 3 | V _ 3 \rangle _ { \rho}
& = &
\langle j _ 1 , m _ 1  | \langle j _ 2 , m _ 2 |
\langle j _ 3 , m _ 3 | V _ 3 \rangle _ { z }
\nonumber \\
& = &
\Big \langle \Psi  _ { j _ 1 , m_ 1 } ( 0 )
\Psi  _ { j _ 2 , m _ 2 } ( 1 )
\Psi _ { j _ 3 , m _ 3 } ( \infty) \Big \rangle
\nonumber \\
& \sim & ( j _ 1 m _ 2 - j _ 2 m _ 1 ) \delta _ { j _ 3 ,
j _ 1 + j + 2 - 1 } \delta _ { m _ 3 ,
m _ 1 + m _ 2 }
\end {eqnarray}
where $Q  _ { j,m} = \oint { dz \over 2 \pi i } \Psi _ { j,m} ( z ) $.
That $w (\infty)$ emerges from the
interaction of discrete states has been stressed in
refs. [12-14]. What we have shown is
that this result easily generalizes to
closed string field theory.

Thus, we have shown that the three-vertex function of string field
theory reproduces the structure constants of $w(\infty)$,
as expected.
Thus, $w(\infty)$ emerges naturally as part of the gauge invariance of
the string field, as in ordinary gauge theory.
(However, the geometric picture of
$w(\infty)$ as area-preserving diffeomorphisms appears to be lost.)

One advantage of this formulation is that the notorious
$ c = 1$ barrier found in matrix models
is easily broken. (However, we do not expect
the model to be exactly soluble beyond this barrier.)

In summary,
we have successfully
formulated 2D string theory as a second quantized
field theory for closed strings. Once Liouville theory is formulated
as a field theory, the mysteries
of 2D string theory (e.g. the field theoretic
origin of $w (\infty)$, discrete
states, etc.) become transparent, because everything is formulated
as a standard gauge theory. See [8] for more details.

\section {Acknowledgments}
We would like to acknowledge partial support from
CUNY-FRAP 6-64435 and NSF PHY-9020495.

\section {References}

\noindent 1.
For a complete review of references, see:
\lq\lq Lectures on 2D Gravity and 2D String Theory,"
by P. Ginsparg and G. Moore, YCTP-P23-92

\noindent 2.
\lq\lq Lecture Notes on 2D Quantum Gravity and
Liouville Theory,"
E. D'Hoker, UCLA/91/TEP/35.

\noindent 3. M. Kaku, {\it Introduction to Superstrings},
Springer-Verlag, New York, 1988.

\noindent 4. M. Kaku, {\it Strings, Conformal Fields, and
Topology}, Springer-Verlag, New York, 1991.

\noindent 5.
M. Kaku, in {\it Functional
Integration, Geometry, and Strings},
Birkhauser Press, Berlin, (1989).

M. Kaku, Phys. Rev. {\bf D41}, 3734 (1990).

\noindent 6. T. Kugo, H. Kunitomo, and K. Suehiro, Phys. Lett. {\bf
226B}, 48 (1989).

\noindent 7. M. Saadi and B. Zwiebach, Ann. Phys. {\bf 192}, 213 (1989).

\noindent 8. M. Kaku, \lq\lq Non-Critical Closed
String Field Theory in $D < 26$ Dimensions,"
CCNY preprint, in preparation, 1993.

\noindent 9. S. Mandelstam, Nucl. Phys.
{\bf B69}, 77 (1974).

\noindent 10. H. Hata, K. Itoh, T. Kugo,
H. Kunitomo and K. Ogawa, Phys. Rev. {\bf D34},
2360 (1986).

\noindent 11. M. Kaku and L. Hua,
Phys. Rev. {\bf D41}, 3748 (1990).

\noindent 12. I.R. Klebanov and A.M. Polyakov,
Mod. Phys. Lett. {\bf A6}, 3273, 1991;

\noindent 13.
E. Witten and B. Zwiebach, Nucl Phys.
{\bf B377}, 55 (1992).

\noindent 14.  I. Ya. Aref' eva, P.B. Medvedev, and A.P.
Zubarev, Steklov Inst. preprint.

\end {document}